\documentclass[aps,prl,superscriptaddress,reprint]{revtex4-1}
\usepackage{amsmath}
\usepackage{amssymb}
\usepackage{graphicx}
\usepackage[colorlinks,urlcolor=blue]{hyperref}
\usepackage{url}
\usepackage{color,xcolor}
\usepackage{ulem}
\usepackage{float}
\usepackage{epstopdf}
\usepackage[none]{hyphenat}
\begin{document}

\title{Electronic structure, self-doping, and superconducting instability in the alternating single-layer trilayer stacking nickelates La$_3$Ni$_2$O$_7$}
\author{Yang Zhang}
\author{Ling-Fang Lin}
\affiliation{Department of Physics and Astronomy, University of Tennessee, Knoxville, Tennessee 37996, USA}
\author{Adriana Moreo}
\affiliation{Department of Physics and Astronomy, University of Tennessee, Knoxville, Tennessee 37996, USA}
\affiliation{Materials Science and Technology Division, Oak Ridge National Laboratory, Oak Ridge, Tennessee 37831, USA}
\author{Thomas A. Maier}
\affiliation{Computational Sciences and Engineering Division, Oak Ridge National Laboratory, Oak Ridge, Tennessee 37831, USA}
\author{Elbio Dagotto}
\affiliation{Department of Physics and Astronomy, University of Tennessee, Knoxville, Tennessee 37996, USA}
\affiliation{Materials Science and Technology Division, Oak Ridge National Laboratory, Oak Ridge, Tennessee 37831, USA}

\date{\today}

\begin{abstract}
Motivated by the recently proposed alternating single-layer trilayer stacking structure for the nickelate La$_3$Ni$_2$O$_7$, we comprehensively study this system using {\it ab initio} and random-phase approximation techniques. Our analysis unveils similarities between this novel La$_3$Ni$_2$O$_7$ structure and other Ruddlesden-Popper nickelate superconductors, such as a similar charge-transfer gap value and orbital-selective behavior of the $e_g$ orbitals.  Pressure primarily increases the bandwidths of the Ni $e_g$ bands, suggesting an enhancement of the itinerant properties of those $e_g$ states. By changing the cell volume ratio $V/V_0$ from 0.9 to 1.10, we found that the bilayer structure in La$_3$Ni$_2$O$_7$ always has lower energy than the single-layer trilayer stacking La$_3$Ni$_2$O$_7$. In addition, we observe a ``self-doping'' effect (compared to the average 1.5 electrons per $e_g$ orbital per site of the entire structure) from the trilayer to the single-layer sublattices and this effect will be enhanced by overall electron doping. Moreover, we find a leading $d_{x^2-y^2}$-wave pairing state that is restricted to the single-layer. Because the effective coupling between the single layers is very weak -- due to the non-superconducting trilayer in between -- this suggests that the superconducting transition temperature $T_c$ in this structure should be much lower than in the bilayer structure.
\end{abstract}

\maketitle
{\it Introduction.--} The Ruddlesden-Popper (RP) perovskite nickelate systems have recently attracted considerable attention due to the discovery of superconductivity under pressure in La$_3$Ni$_2$O$_7$ (327-LNO) with a $d^{7.5}$ configuration~\cite{Sun:arxiv}, opening a remarkable and challenging avenue for the study of nickelate-based unconventional high-{\it $T_{\rm c}$} superconductivity~\cite{LiuZhe:arxiv,Zhang:arxiv-exp,Hou:arxiv,Yang:arxiv09,Zhang:arxiv09,Wang:arxiv9}.

\begin{figure*}
\centering
\includegraphics[width=0.84\textwidth]{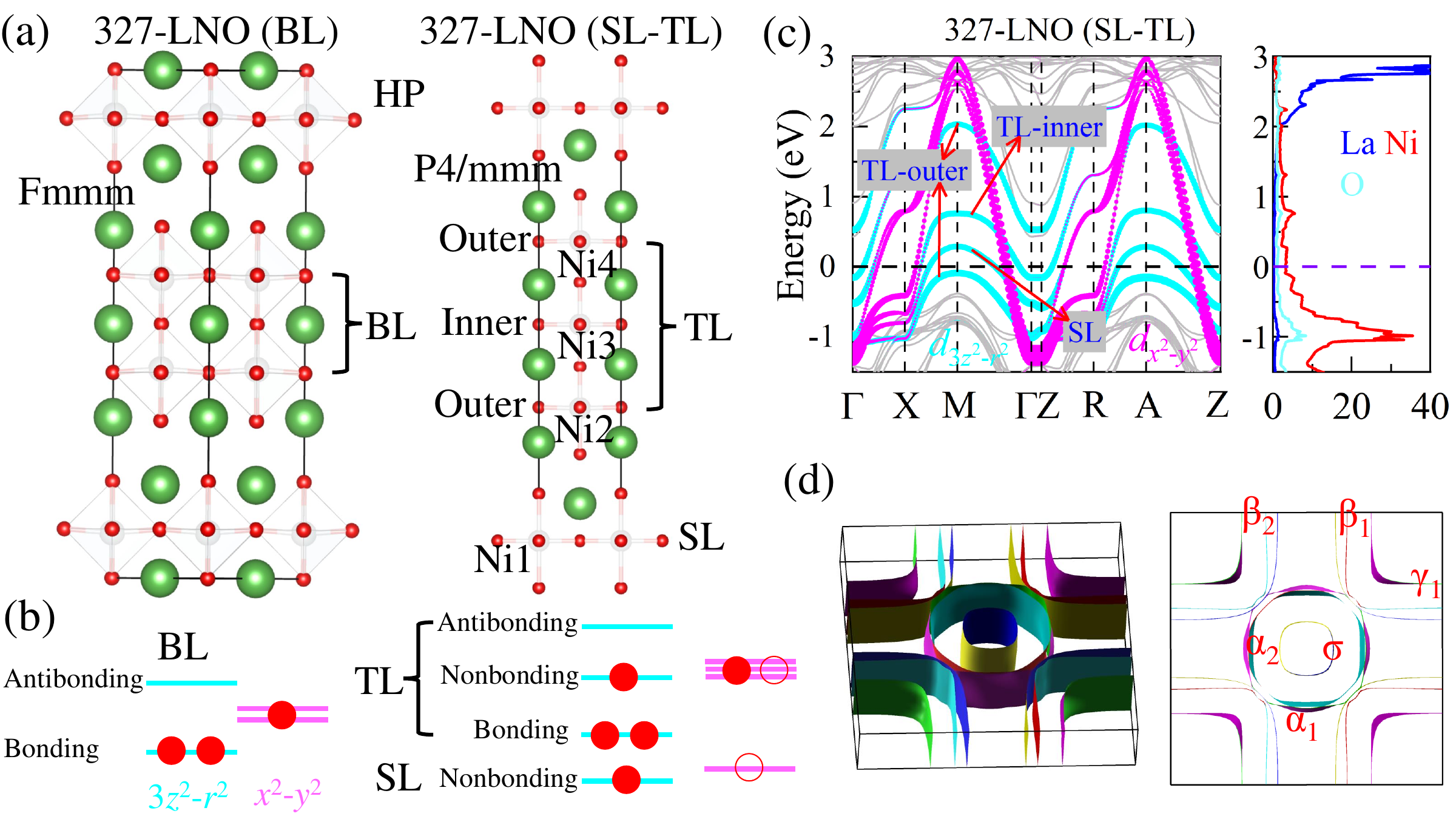}
\caption{(a) Schematic crystal structure of the conventional cells of the high-pressure Fmmm phase of 327-LNO (BL) and P4/mmm phase of 327-LNO (SL-TL), respectively (green = La; gray = Ni; red = O). All crystal structures were visualized using the VESTA code~\cite{Momma:vesta}. (b) Sketches of electronic states in the BL and SL-TL.
The light blue (pink) horizontal lines represent $d_{3z^2-r^2}$ ($d_{x^2-y^2}$) states.
The solid and open circles represent 1.0 and 0.5 electrons, respectively.
The total population of $e_g$ electrons considered is $n = 3.0$ and $n=6.0$ electrons, for BL with two sites and SL-TL with four sites, respectively. (c) The Ni $e_g$-orbitals projected band structures and density of states. The $d_{3z^2-r^2}$ and $d_{x^2-y^2}$ orbitals are distinguished by the blue and red lines. (d) FS for the non-magnetic state of the high-pressure P4/mmm phase of the 327-LNO (SL-TL) structure at 16 GPa. Note that the local $z$-axis is perpendicular to the NiO$_6$ plane towards the top O atom, while the local $x$- or $y$-axis is along the in-plane Ni-O bond directions.}
\label{Crystal}
\end{figure*}

For the 327-LNO system, the original experiments reported a bilayer (BL) stacking structure~\cite{Sun:arxiv}, where superconductivity was found in a very broad pressure range, with a high transition temperature up to 80 K~\cite{Sun:arxiv}. This occurs after a first-order structural transition from the ambient-pressure Amam phase with distorted NiO$_6$ octahedra to the high-pressure Fmmm phase without tilting of those oxygen octahedra (see Fig.~\ref{Crystal}(a)). Very recently, a transition to a tetragonal I4/mmm phase under pressure has also been proposed in both theory~\cite{Geisler:qm} and experiment~\cite{Wang:jacs}. Because the distortion from I4/mmm (No. 139) to Fmmm (No. 69) is very small, there is no fundamental differences among those two phases, providing the same physics~\cite{Sakakibara:prl24}. Based on the BL structure, many follow-up experimental studies~\cite{Kakoi:arxiv12,Dong:arxiv12,Xie:arxiv2024,Chen:arxiv2024,Dan:arxiv2024,Takegami:prb} and theoretical~\cite{Luo:prl23,Zhang:prb23,Christiansson:prl23,Yang:prb23,Sakakibara:prl24,Shen:cpl,Liu:prl23,Zhang:arxiv1,Yang:prb,Oh:prb23,
Liao:prb23,Cao:prb23,Lechermann:prb23,Shilenko:prb23,Jiang:cpl24,Huang:prb23,Qin:prb23,Zhang:prb23-2,Zhang:prb24,Kaneko:prb24,Qu:prl,Lu:prl} efforts based on two ``active'' orbitals ($d_{3z^2-r^2}$ and $d_{x^2-y^2}$) were presented to understand the interesting physical properties and possible superconducting pairing mechanism in this compound. Very recently, J. Li {\it et.~al.}~\cite{Li:arxiv24} observed the Meissner effect of the superconducting state using the $ac$ magnetic susceptibility, with the superconducting volume fraction being around $50 \%$.

Furthermore, an interesting alternating single-layer (SL) -- trilayer (TL) stacking structure was proposed experimentally (with the same chemical formula 327-LNO) as an alternative to the BL stacking structure (see Fig.~\ref{Crystal}(a)), by several groups~\cite{Chen:jacs,Puphal:arxiv12,Wang:ic,Abadi:arxiv24}. Interestingly, signatures of superconductivity under pressure were also reported in another RP  nickelate La$_4$Ni$_3$O$_{10}$ with only a trilayer stacking structure~\cite{Zhang:nc20,Samarakoon:prx,Sakakibara:arxiv09,Li:cpl,Zhu:arxiv11,Zhang:arxiv11,Li:arxiv11}. However, superconductivity was absent in SL La$_2$NiO$_4$ under pressure~\cite{Zhang:arxiv09}.

Similarly to 327-LNO (BL), the corner-sharing NiO$_6$ octahedra were also found to be strongly distorted in TL La$_4$Ni$_3$O$_{10}$ at ambient conditions~\cite{Zhang:nc20,Samarakoon:prx,Sakakibara:arxiv09,Li:cpl,Zhang:arxiv11,Zhu:arxiv11,Zhang:arxiv11,Li:arxiv11}, leading to a monoclinic P2$_1$/c structure (No. 14). Increasing pressure suppresses the distortion of the NiO$_6$ octahedra, yielding a high-symmetry I4/mmm phase without the tilting of oxygen octahedra around 15 GPa, where the signature of superconductivity was found with a $T_c$ of about $20-30$ K~\cite{Sakakibara:arxiv09,Li:cpl,Zhu:arxiv11,Zhang:arxiv11,Li:arxiv11}. In 327-LNO (SL-TL), the NiO$_6$ oxygen octahedra also do not display any substantial tilting in the high-pressure P4/mmm phases~\cite{Chen:jacs,Puphal:arxiv12,Wang:ic,Lechermann:1313}.

Considering these developments, several interesting questions naturally arise: What are the similarities and differences between the 327-LNO (SL-TL) and other RP nickelate superconductors under pressure? What interesting physics can be obtained in this alternating SL and TL stacking 327-LNO, as compared to the first proposed bilayer 327-LNO structure? Could we obtain superconductivity in 327-LNO (SL-TL) and is the pairing mechanism the same as in other RP nickelate superconductors?

{\it Electronic structure of the P4/mmm phase of 327-LNO (SL-TL)--} In the 327-LNO (BL) system, due to the ``dimer'' physics induced by its bilayer geometry~\cite{Zhang:prb23}, the $d_{3z^2-r^2}$ orbital splits into antibonding and bonding states, while the $d_{x^2-y^2}$ orbital remains decoupled among planes, as shown in Fig.~\ref{Crystal}(b), resulting in an orbital-selective behavior~\cite{Streltsovt:prb14,Zhang:ossp}. Considering the nominal valence  Ni$^{2.5+}$, close to two electrons would occupy the bonding states of the $d_{3z^2-r^2}$ orbital and approximately one electron stays in the $d_{x^2-y^2}$ band~\cite{Zhang:prb23} (see Fig.~\ref{Crystal}(b)).

Because 327-LNO (SL-TL) has an alternating SL and TL stacking structure, the $d_{3z^2-r^2}$ orbital from the TL sublattice also splits into antibonding, nonbonding, and bonding states due to the ``trimer'' physics induced by the TL structure~\cite{Zhang:arxiv24}, as displayed in Fig.~\ref{Crystal}(b). Meanwhile, the $d_{x^2-y^2}$ orbital does not participate in the formation of antibonding and bonding states, as well as the $e_g$ states from the SL sublattice. Considering an average of 1.5 electrons in $e_g$ states per Ni in the 327-LNO,  two electrons would occupy the bonding states of the $d_{3z^2-r^2}$ orbital and one electron enters the nonbonding state in the TL sublattice, while an extra 1.5 electrons would remain in the $d_{x^2-y^2}$ state [Fig.~\ref{Crystal}(b)]. For the SL sublattice, 1.0 and 0.5 electrons would occupy the states of the $d_{3z^2-r^2}$ and $d_{x^2-y^2}$ orbitals, respectively. Being a SL, these are, obviously, only nonbonding states. Then, the $e_g$ orbitals of 327-LNO (SL-TL) have both orbital-selective and layer-selective characteristics.

To better understand the physical properties of 327-LNO (SL-TL), we have calculated the electronic structure based on the experimental crystal structure under pressure~\cite{HP} using first-principles density functional theory (DFT)~\cite{Kresse:Prb,Kresse:Prb96,Blochl:Prb,Perdew:Prl}. The states near the Fermi level  mainly arise from Ni $3d$ orbitals hybridized with O $p$-states, which are located at lower energies than the Ni $3d$ states, indicating a charge-transfer picture similar to other nickelates~\cite{Zhang:prb20,Zhang:prb23,Zhang:prb23-2}. As expected, the $d_{3z^2-r^2}$ orbital from the TL sublattice splits into three antibonding, nonbonding, and bonding related bands in the 327-LNO (SL-TL) band structure, as shown in Fig.~\ref{Crystal}(c). In addition, the $d_{x^2-y^2}$ orbital from the TL sublattice and the $e_g$ orbitals from the SL sublattice all remain nearly decoupled between different planes.

Similarly as it occurs in 327-LNO (BL), due to the strong in-plane hybridization of the $e_g$ states~\cite{Zhang:prb23} electrons transfer between the $d_{3z^2-r^2}$ and $d_{x^2-y^2}$ orbitals in 327-LNO (SL-TL) as well, leading to a noninteger electronic population in both orbitals. A crucial observation is that different from the TL La$_4$Ni$_3$O$_{10}$, the bonding band of the $d_{3z^2-r^2}$ states from the TL sublattice {\it does not touch} the Fermi surface (FS). Therefore, the TL states do not create a Fermi surface $\gamma$ pocket in 327-LNO (SL-TL) (see Fig.~\ref{Crystal}(d)). Instead, a $\gamma$ pocket does exist in the TL La$_4$Ni$_3$O$_{10}$ context~\cite{Zhang:arxiv24}), basically due a different crystal field. Note that the $\gamma_1$ sheet shown in Fig.~\ref{Crystal}(d) arises {\it entirely} from the SL sublattice, and should not be confused with the $\gamma$ pocket of the BL and TL systems. The TL sublattice states give rise to two $\beta$ sheets and an $\alpha_1$ sheet of mixed $e_g$ orbital character, while the $\sigma$ pocket is made up of the $d_{3z^2-r^2}$ orbital in the 327-LNO (SL-TL), similar to La$_4$Ni$_3$O$_{10}$. Moreover, the SL sublattice also gives rise to the $\alpha_{2}$ sheet displayed in Fig.~\ref{Crystal}(d).

\begin{figure}
\centering
\includegraphics[width=0.48\textwidth]{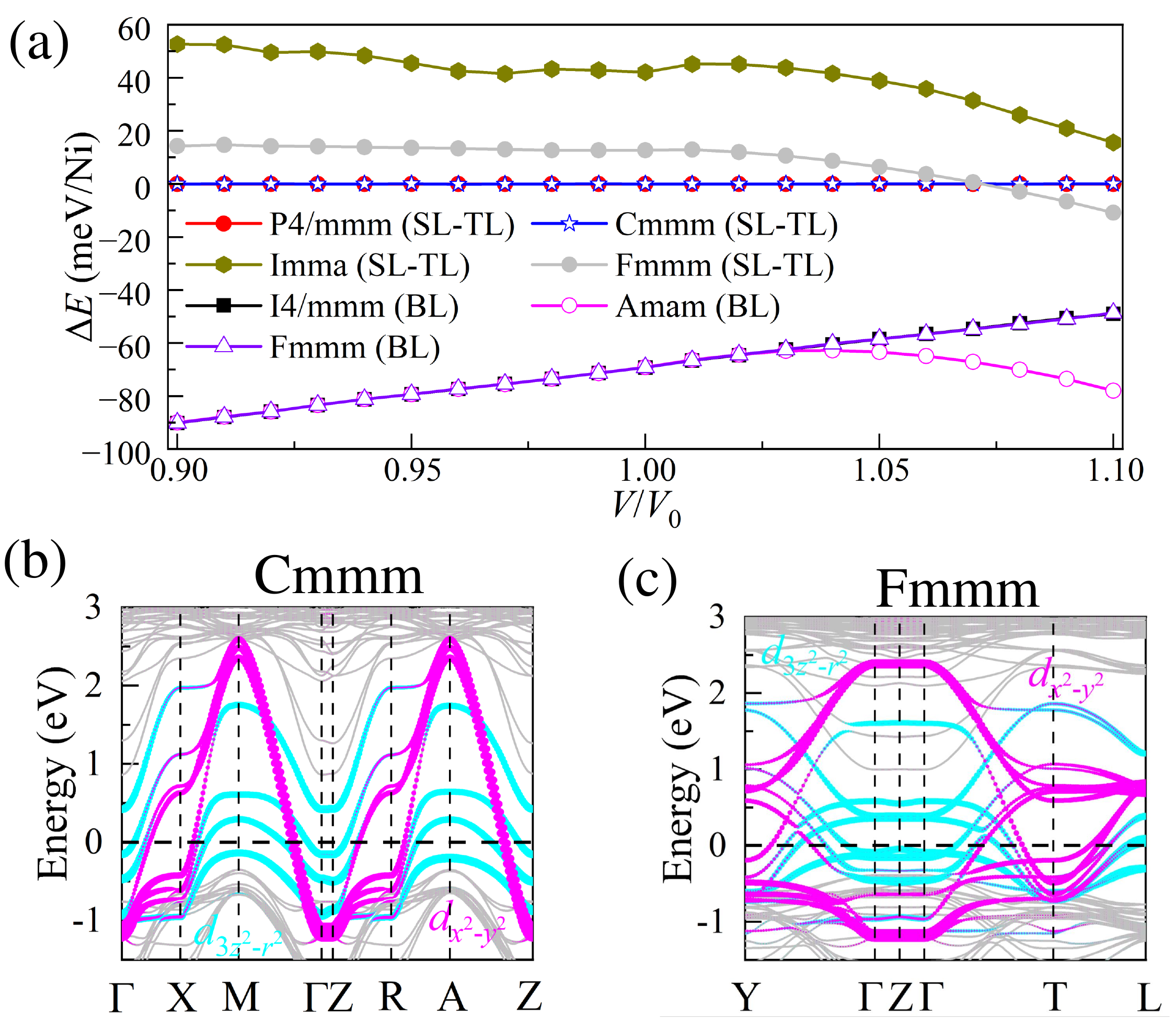}
\caption{(a) Calculated energies of different structural phases of 327-LNO (SL-TL) and 327-LNO (BL), as a function of the ratio $V/V_0$. Here, $V_0$ is the conventional-cell volume of the P4/mmm phase of 327-LNO (SL-TL) at 16 GPa. The P4/mmm phase of 327-LNO (SL-TL) is taken as the energy of reference. (b-c) The Ni $e_g$-orbitals projected band structure for the (b) Cmmm and (c) Fmmm phases, respectively.}
\label{V-E-band}
\end{figure}

{\it Pressure effect --} Furthermore, the Imma~\cite{Puphal:arxiv12} and Fmmm~\cite{Puphal:arxiv12,Lechermann:1313}, with octahedral tilting distortion, were also proposed as possible structures at ambient pressure. By fully relaxing these phases, we found that the Fmmm and P4/mmm phases have the lowest energy or enthalpy at 0 and 16 GPa, respectively. To understand better the different phases of 327-LNO (SL-TL) and 327-LNO (BL), we also studied the energies when changing the ratio $V/V_0$, where $V_0$ is the conventional-cell volume of the P4/mmm phase of 327-LNO (SL-TL) at 16 GPa. As shown in Fig.~\ref{V-E-band}(a), 327-LNO (BL) always has lower energy than 327-LNO (SL-TL) when changing the ratio $V/V_0$ from 0.9 to 1.10~\cite{U4}. Because the distortion from I4/mmm  to Fmmm in 327-LNO (BL) is very small, they almost have identical energies, as displayed in Fig.~\ref{V-E-band}(a).

In addition, as the volume increases, corresponding to a decrease in pressure, the Amam phase has lower energy than the Fmmm phase in the 327-LNO (BL) structure, in agreement with a previous experiment~\cite{Sun:arxiv}. For 327-LNO (SL-TL), the P4/mmm and Cmmm phases have nearly degenerate energies in the entire $V/V_0$ region that we studied because the distortions from the high symmetry P4/mmm phase are quite small. In addition, at very large $V/V_0$, the Fmmm phase has the lowest energy among those phases, while the P4/mmm phase has lower energy than that in the Fmmm and Imma phases by reducing the ratio  $V/V_0$.

Furthermore, we also calculated the band structure of the Cmmm and Fmmm phases of 327-LNO (SL-TL) by using the primitive unit cell at ambient pressure, as displayed in Figs.~\ref{V-E-band}(b) and (c). Similarly to the P4/mmm phase under pressure, the bonding state of TL $d_{3z^2-r^2}$ orbitals does not touch the Fermi level. Thus, this band does not produce a hole $\gamma$ pocket in the FS at ambient pressure, which is in agreement with recent angle-resolved photoemission spectroscopy experiments~\cite{Abadi:arxiv24}.

{\it Tight-binding model and self-doping effect --} To understand the low-energy physics under pressure, we constructed an eight-band ${e_g}$-orbital tight-binding model for the SL and TL sublattice states for the P4/mmmm phase of 327-LNO (SL-TL) at 16 GPa, with overall filling $n = 6$, including the longer-range hoppings between the SL and TL sublattices. The entire hopping file used in the present work can be found in Supplementary Material~\cite{Supplemental}, where additional details are provided for the tight-binding model.

\begin{figure}
\centering
\includegraphics[width=0.42\textwidth]{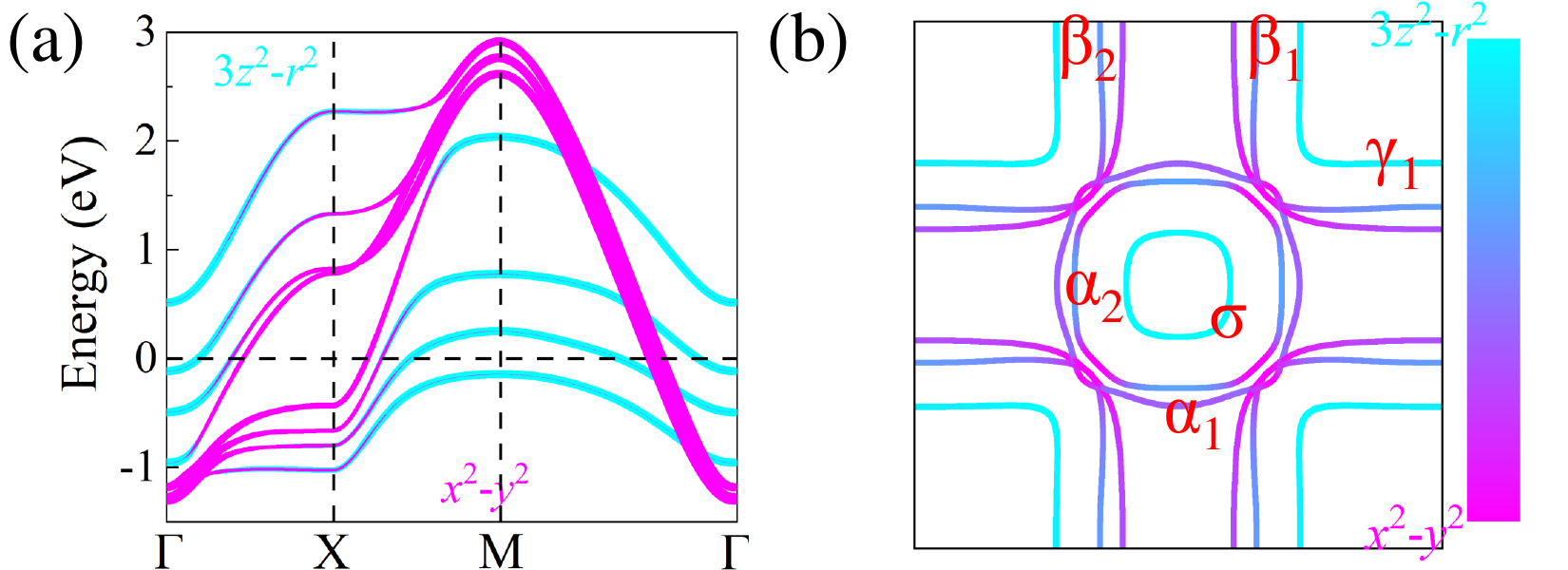}
\includegraphics[width=0.42\textwidth]{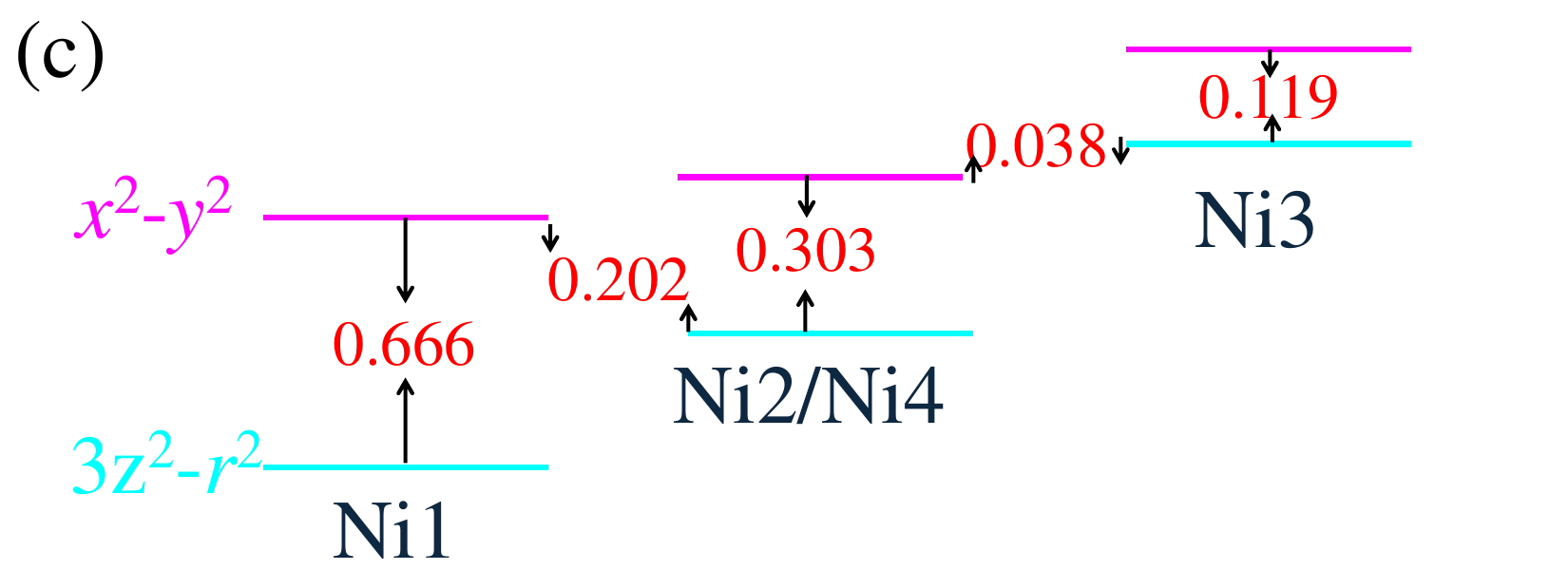}
\includegraphics[width=0.42\textwidth]{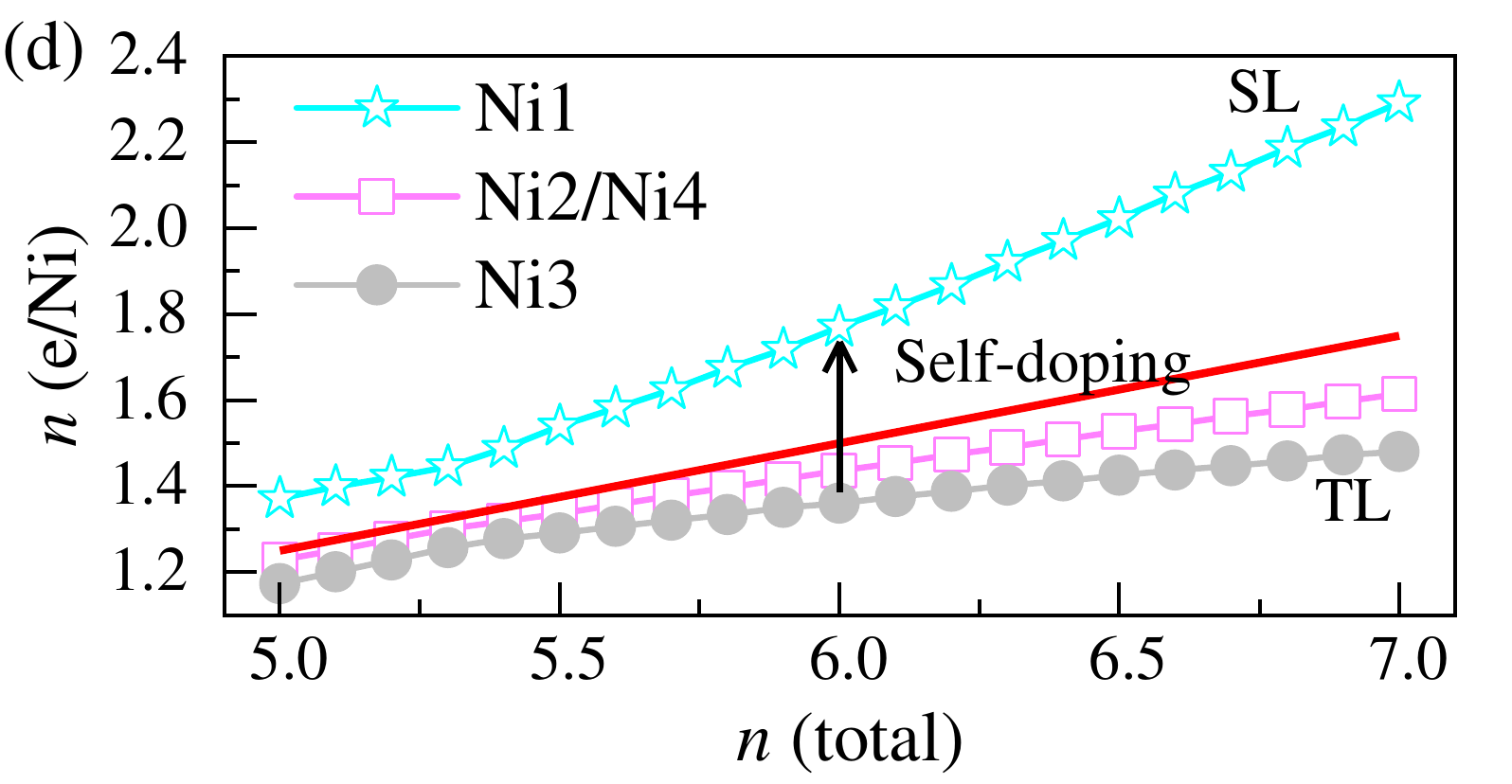}
\includegraphics[width=0.42\textwidth]{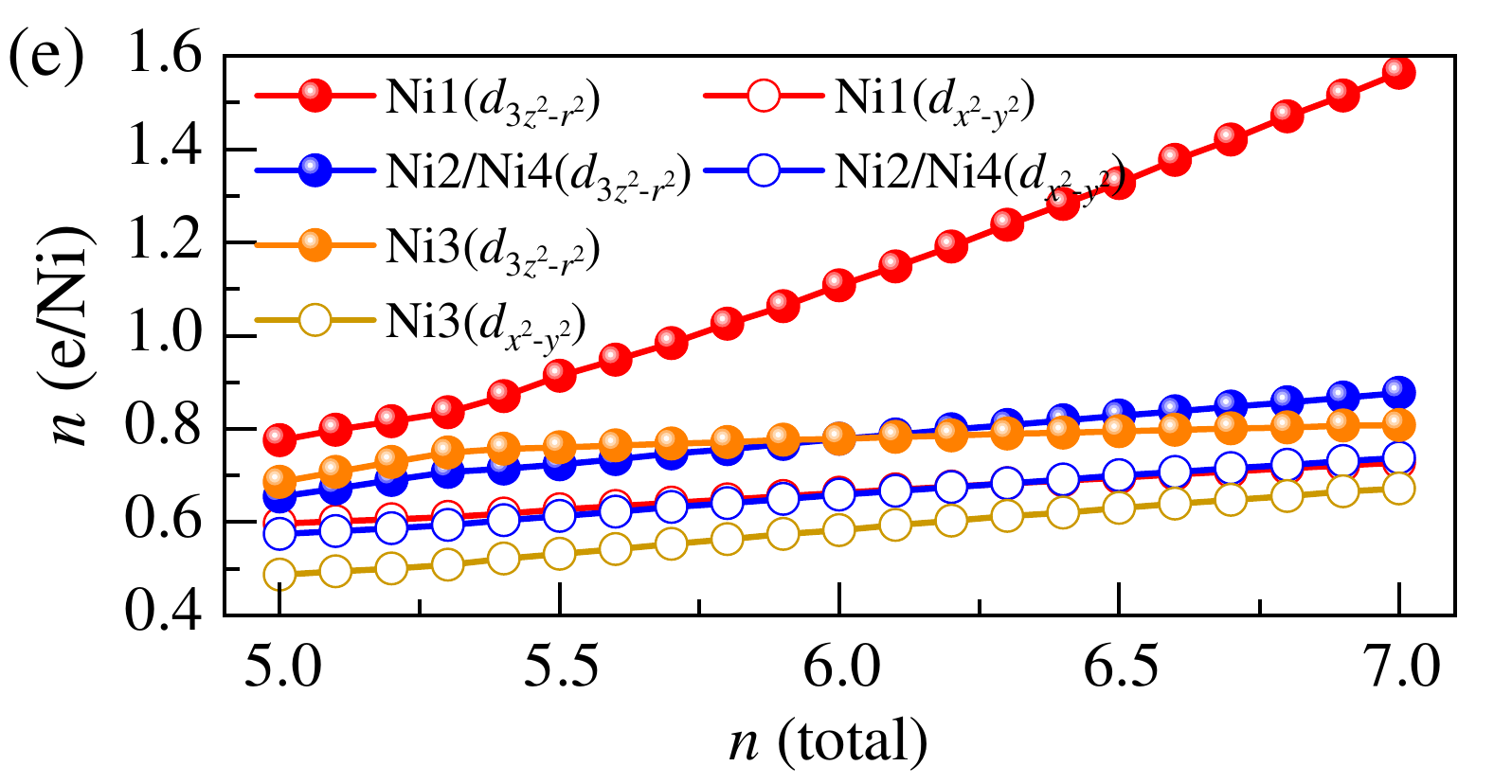}
\caption{(a) Tight-binding band structure and (b) Fermi surface for the P4/mmm phase of 327-LNO (SL-TL) at 16 GPa. Here, an eight-band $e_g$ orbital tight-binding model was considered including SL and TL hoppings (see the hopping file in the Supplementary Material~\cite{Supplemental}) with an overall filling of $n = 6$ (i.e. 1.5 electrons per site). (c) Crude sketches of the crystal-field splitting of the $e_g$ orbitals for different Ni sites. All the values are given in units of eV. (d) The total electron occupations and (e) different electronic densities of the $e_g$ orbitals for the different Ni sites vs. the overall filling $n$ in the tight-binding model. The red line in panel (d) represents the average electronic Ni site occupancy.}
\label{TB}
\end{figure}

As shown in Figs.~\ref{TB}(a) and (b), the tight-binding band structure and FS fit very well with the DFT results discussed in the previous section. By using the maximally localized Wannier functions (MLWFs) method based on the WANNIER90 package~\cite{Mostofi:cpc}, we obtained the crystal-field splitting of the $e_g$ orbitals for different Ni sites in the P4/mmm phase of 327-LNO (SL-TL), as summarized in Fig.~\ref{TB}(c). The Ni1 site from the SL sublattice has a larger crystal-field splitting ($\Delta \sim 0.666$ eV) than the Ni2, Ni3, and Ni4 sites from the TL sublattice. In the TL sublattice, the outer layers (Ni2 and Ni4 sites) have lower on-site energies than the middle layer (Ni3 site), similarly to the case of the TL for La$_4$Ni$_3$O$_{10}$. Furthermore, we find that the ratios $t^{12}/t^{22}$  are 0.425, 0.511, and 0.548 for the Ni1, Ni2/Ni4 and Ni3 sites, respectively (where the in-plane inter-orbital hopping between the $d_{3z^2-r^2}$ and $d_{x^2-y^2}$ orbitals is $t^{12}$, and the intra-orbital  $d_{x^2-y^2}$ hopping is $t^{22}$). This  indicates that the in-plane hybridization of the $e_g$ orbitals is stronger in the TL sublattice.

Considering the average electronic density of the $e_g$ states (1.5 per Ni) in 327-LNO (SL-TL), the calculated electronic densities are approximately 1.770, 1.436, and 1.360 for Ni1, Ni2/Ni4 and Ni3 sites, respectively. This indicates a ``self-doping'' effect, where electrons are transferred from the TL to the SL sublattice. Note that here the sum of the electronic populations of Ni2, Ni3, and Ni4 ($\sim 4.23$), locates the system in the superconducting region of La$_4$Ni$_3$O$_{10}$, where $d_{3z^2-r^2}$ $\gamma$ pockets are obtained according to Ref.~\cite{Zhang:arxiv24}. The primary intuitive reason why TL La$_4$Ni$_3$O$_{10}$ has a $\gamma$ pocket, and thus superconductivity, but TL from 327-LNO (SL-TL) does not, is because of the crystal-field value: it is larger in TL from 327-LNO (SL-TL) than in TL La$_4$Ni$_3$O$_{10}$.

Moreover, the outer Ni layers (Ni2 and Ni4 sites) of the TL sublattice have more electrons than the middle layer (Ni3 site). In addition, we also calculated the electron occupations for different Ni sites by changing the overall filling $n$ from 5.0 to 7.0, as displayed in Figs.~\ref{TB}(d) and (e). For electronic doping, extra electrons move to the SL Ni1 site with an increased difference in electron densities between the SL and TL sublattices, suggesting an enhanced ``self-doping'' effect. Specifically, in the hole-doping region, the SL Ni1 site loses more electrons than other sites in the TL sublattice. As shown in Fig.~\ref{TB}(e), for the doping region we studied, the electronic occupation of the $d_{3z^2-r^2}$ orbitals is larger than that in the $d_{x^2-y^2}$ orbital.

\begin{figure}
\centering
\includegraphics[width=0.46\textwidth]{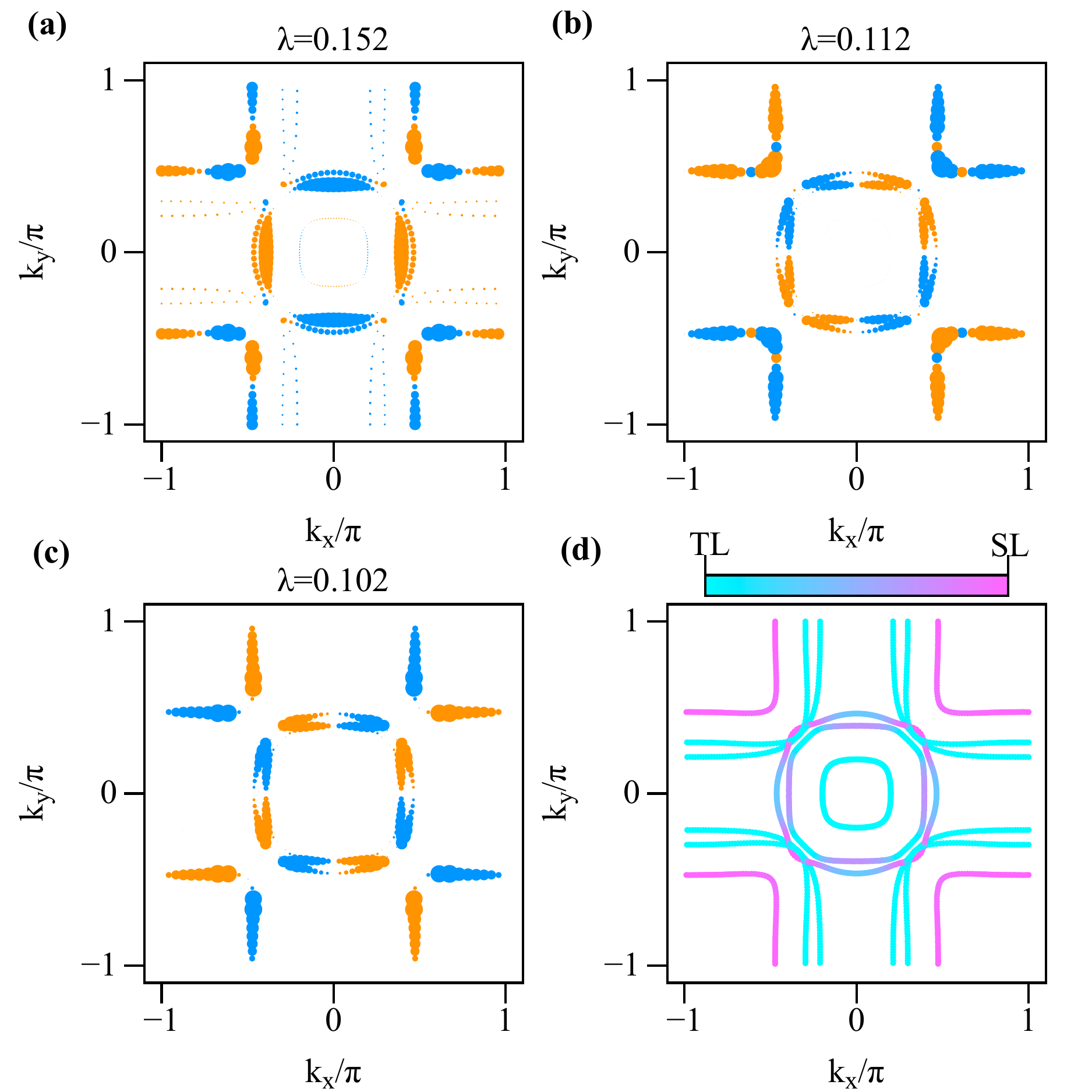}
\caption{The RPA calculated leading superconducting singlet gap structure $g_\alpha({\bf k})$ for momenta ${\bf k}$ on the Fermi surface of 327-LNO (SL-TL) at $n = 6.0$ and their pairing strength $\lambda$: (a) Leading $d_{x^2-y^2}$-wave state with $\lambda=0.152$, (b) subleading $d_{xy}$-wave state with $\lambda=0.112$, (c) $g$-wave state with $\lambda=0.102$. The sign of the gap structure $g_\alpha({\bf k})$ is indicated by the colors (orange = positive, blue = negative), and the gap amplitude by the point size. The RPA calculations used Coulomb interaction parameters $U = 0.5$ (intra-orbital), $U'=U/2$ (inter-orbital), and $J = J' = U/4$ (Hund coupling and pair hopping, respectively) in units of eV. The hopping parameters are available in the Supplementary Material~\cite{Supplemental}. The dominant character of the Fermi surface Bloch states arising from the SL (magenta) and TL (cyan) sublattices is shown in panel (d).}
\label{Pairing}
\end{figure}

{\it RPA pairing tendencies --} To analyze the superconducting pairing tendencies, we used multi-orbital random phase approximation (RPA) calculations to assess the tight-binding model including the SL and TL sublattices. The RPA calculations of the pairing vertex are based on a perturbative weak-coupling expansion in the local Coulomb interaction matrix, which contains intra-orbital ($U$), inter-orbital ($U'$), Hund's rule coupling ($J$), and pair hopping ($J'$) terms ~\cite{Kubo2007,Graser2009,Altmeyer2016,Romer2020}. The pairing strength $\lambda_\alpha$ for pairing channel $\alpha$ and its corresponding pairing structure $g_\alpha({\bf k})$ are obtained by solving an eigenvalue problem of the form
\begin{eqnarray}
\int_{FS} d{\bf k'} \, \Gamma({\bf k -  k'}) g_\alpha({\bf k'}) = \lambda_\alpha g_\alpha({\bf k}),\,
\label{eq:pp}
\end{eqnarray}
where the momenta ${\bf k}$ and ${\bf k'}$ are restricted to the FS, and $\Gamma({\bf k - k'})$ is the irreducible particle-particle vertex. In the RPA approximation, the dominant term entering $\Gamma({\bf k-k'})$ is the RPA spin susceptibility $\chi_s({\bf k-k'})$. Here, the parameters in the kinetic hopping component of the Hamiltonian are taken from the P4/mmmm phase of 327-LNO (SL-TL) at 16 GPa.

By solving the eigenvalue problem in Eq.~\ref{eq:pp} for the RPA calculated pairing vertex $\Gamma({\bf k-k'})$ of 327-LNO (SL-TL) at $n = 6.0$, we find that the $d_{x^2-y^2}$ channel is the leading pairing instability caused by spin fluctuations, followed by subleading $d_{xy}$- and $g$-wave solutions, as displayed in Figs.~\ref{Pairing}(a-c). Here we used $U=0.5$ eV, $U'=U/2$, $J=J'=U/4$. Despite the small value of $U$, the pairing strength $\lambda=0.152$ of the leading $d_{x^2-y^2}$ chanel is already significant. As a comparison, for the bilayer 327 structure, we obtained $\lambda \sim 0.2$ at 15 GPa for $U=0.8$ eV~\cite{Zhang:arxiv1}. Note that {\it both} the $d_{x^2-y^2}$ and $d_{3z^2-r^2}$ orbitals have contributions to the superconducting pairing channels ($d_{x^2-y^2}$-, $d_{xy}$- and $g$-wave), similarly to previous discussion in BL 327-LNO studies~\cite{Zhang:arxiv1,Tian:prb24}.

Moreover, we note that for all three leading pairing states, the pairing gap is only finite on the regions of the Fermi surface made up of states from the SL sublattice (magenta regions in panel (d) in Fig.~\ref{Pairing}), and negligible in the TL dominated regions. This suggests that pairing, while strong, is restricted to the SL sublattice, and caused by strong intra-SL-sublattice spin fluctuations between the $\alpha_2$ and $\gamma_1$ Fermi surface sheets. We believe that it is the absence of a TL $\gamma$-pocket that pairing in the TL sublattice is weak. Based on these results, we speculate that a possible three-dimensional superconducting transition involving the SL system would be suppressed to very low temperatures, because the effective coupling between different SL sublattices is negligible due to the presence of the TL in between. In the experimental doping region $-15 \%$ to $15 \%$, corresponding to $n = 5.4$ to $6.6$ in our calculation, we do not observe the hole $\gamma$ pocket from the bonding state of the TL $d_{3z^2-r^2}$ orbital arising at the FS. Thus, from our perspective, superconductivity is difficult to stabilize in the 327-LNO (SL-TL) nickelate system even though a superconducting instability still can survive in the SL lattice under doping. For the SL alone, such as in  the La$_2$NiO$_4$ system, it is very interesting to investigate the possible superconductivity under doping, which needs and deserves further detailed studies. However, this is outside the scope of our present work.

In addition, we find that larger $U$ ($U \textgreater 0.6$ eV) causes a spin-density wave instability in the SL sublattice with ${\bf q}\sim (\pi, \pi, 0)$, corresponding to the conventional N\'eel G-type antiferromagnetic order, arising from the near perfect nesting of the SL $\alpha_2$ and $\gamma_1$ sheet Fermi surface states (magenta regions in Fig.~\ref{Pairing}(d)).

{\it Conclusions.--} In summary, we presented a systematic study of the newly reported 327-LNO (SL-TL), with an alternating single-layer and trilayer stacking structure, by using DFT and RPA calculations. We found a large charge-transfer energy, a common character with other nickelates.  By changing the volume ratio $V/V_0$ from 0.9 to 1.10, corresponding to the effect of pressure, we found that 327-LNO (BL) always has lower energy than 327-LNO (SL-TL). Moreover, we observed a ``self-doping'' effect from the TL to the SL sublattices, compared to the average 1.5 electrons $e_g$ orbital per site of the entire structure, which is enhanced by electron doping. Based on RPA calculations, we obtained a leading $d_{x^2-y^2}$-wave pairing state with sizable pairing strength $\lambda$. However, this pairing state is restricted to the SL sublattice and, therefore, a three-dimensional superconducting state will be suppressed to very low temperatures due to the weak coupling between the SL sublattices, which will reduce the coherence between those layers. These results suggest that a possible superconducting $T_c$ in 327-LNO (SL-TL) may not be sufficiently high to match the experimentally observed $T_c$ in 327-LNO. Consequently, we believe that a bilayer stacking structure remains the most likely in 327-LNO with $T_c \sim 80$ K.

This work was supported by the U.S. Department of Energy (DOE), Office of Science, Basic Energy Sciences (BES), Materials Sciences and Engineering Division.


\begin{references}
\bibitem{Sun:arxiv} H. Sun, M. Huo, X. Hu, J. Li, Y. Han, L. Tang, Z. Mao, P. Yang, B. Wang, J. Cheng, D.-X. Yao, G.-M. Zhang, and M. Wang, Signatures of superconductivity near 80 K in a nickelate under high pressure \href{https://doi.org/10.1038/s41586-023-06408-7}{Nature \textbf{621} 493 (2023).}
\bibitem{LiuZhe:arxiv} Z. Liu, M. Huo, J. Li, Q. Li, Y. Liu, Y. Dai, X. Zhou, J. Hao, Y. Lu, M. Wang, and W.-H. Wen,  Electronic correlations and partial gap in the bilayer nickelate La$_3$Ni$_2$O$_7$ \href{https://doi.org/10.48550/arXiv.2307.02950}{arXiv 2307.02950 (2023).}
\bibitem{Zhang:arxiv-exp} Y. Zhang, D. Su, Y. Huang, H. Sun, M. Huo, Z. Shan, K. Ye, Z. Yang, R. Li, M. Smidman, M. Wang, L. Jiao, and H. Yuan, High-temperature superconductivity with zero-resistance and strange metal behavior in La$_3$Ni$_2$O$_{7-\delta}$ \href{https://doi.org/10.1038/s41567-024-02515-y}{Nat. Phys. (2024).}
\bibitem{Hou:arxiv} J. Hou, P. T. Yang, Z. Y. Liu, J. Y. Li, P. F. Shan, L. Ma, G. Wang, N. N. Wang, H. Z. Guo, J. P. Sun, Y. Uwatoko, M. Wang, G.-M. Zhang, B. S. Wang, and J.-G. Cheng, Emergence of High-Temperature Superconducting Phase in Pressurized La$_3$Ni$_2$O$_7$ Crystals \href{https://doi.org/10.48550/arXiv.2307.09865}{Chin. Phys. Lett. \textbf{40} 117302 (2023).}
\bibitem{Yang:arxiv09} J. Yang, H. Sun, X. Hu, Y. Xie, T. Miao, H. Luo, H. Chen, B. Liang, W. Zhu, G. Qu, C.-Q. Chen, M. Huo, Y. Huang, S. Zhang, F. Zhang, F. Yang, Z. Wang, Q. Peng, H. Mao, G. Liu, Z. Xu, T. Qian, D.-X. Yao, M. Wang, L. Zhao, and X. J. Zhou, Orbital-dependent electron correlation in double-layer nickelate La$_3$Ni$_2$O$_7$ \href{https://doi.org/10.1038/s41467-024-48701-7}{Nat. Commun.  \textbf{15} 4373 (2024).}
\bibitem{Zhang:arxiv09} M. Zhang, C. Pei, Q. Wang, Y. Zhao, C. Li, W. Cao, S. Zhu, J. Wu, and Y. Qi, Effects of pressure and doping on Ruddlesden-Popper phases La$_{\rm n+1}$Ni$_n$O$_{\rm 3n+1}$ \href{https://doi.org/10.1016/j.jmst.2023.11.011}{J. Mater. Sci. Technol.  \textbf{185} 147 (2024).}
\bibitem{Wang:arxiv9} G. Wang, N. N. Wang, X. L. Shen, J. Hou, L. Ma, L. F. Shi, Z. A. Ren, Y. D. Gu, H. M. Ma, P. T. Yang, Z. Y. Liu, H. Z. Guo, J. P. Sun, G. M. Zhang, S. Calder, J.-Q. Yan, B. S. Wang, Y. Uwatoko, and J.-G. Cheng, Pressure-Induced Superconductivity In Polycrystalline La$_3$Ni$_2$O$_{7-\delta}$ \href{https://doi.org/10.1103/PhysRevX.14.011040}{Phys. Rev. X \textbf{14} 011040 (2024).}
\bibitem{Geisler:qm} B. Geisler, J. J. Hamlin, G. R. Stewart, R. G. Hennig and P. J. Hirschfeld, Structural transitions, octahedral rotations, and electronic properties of $A_3$Ni$_2$O$_7$ rare-earth nickelates under high pressure \href{https://doi.org/10.1038/s41535-024-00648-0}{npj Quantum Mater. \textbf{9} 38 (2024).}
\bibitem{Wang:jacs} L. Wang, L. Wang, Y. Li, S.-Y. Xie, F. Liu, H. Sun, C. Huang, Y. Gao, T. Nakagawa, B. Fu, B. Dong, Z. Cao, R. Yu, S. I. Kawaguchi, H. Kadobayashi, M. Wang, C. Jin, H.-k. Mao, and H. Liu, Structure Responsible for the Superconducting State in La$_3$Ni$_2$O$_7$ at High-Pressure and Low-Temperature Conditions \href{https://doi.org/10.1021/jacs.3c13094}{J. Am. Chem. Soc. \textbf{1} 7506 (2024).}
\bibitem{Sakakibara:prl24} H. Sakakibara, N. Kitamine, M. Ochi, and K. Kuroki, Possible High $T_c$ Superconductivity in La$_3$Ni$_2$O$_7$ under High Pressure through Manifestation of a Nearly Half-Filled Bilayer Hubbard Model \href{https://doi.org/10.1103/PhysRevLett.132.106002}{Phys. Rev. Lett. \textbf{132} 106002 (2024).}
\bibitem{Kakoi:arxiv12} M. Kakoi, T. Oi, Y. Ohshita, M. Yashima, K. Kuroki, Y. Adachi, N. Hatada, T. Uda, H. Mukuda, Multiband Metallic Ground State in Multilayered Nickelates La$_3$Ni$_2$O$_7$ and La$_4$Ni$_3$O$_{10}$ Probed by $^{139}$La-NMR at Ambient Pressure \href{https://doi.org/10.48550/arXiv.2312.11844}{J. Phys. Soc. Jpn. \textbf{93} 053702 (2024).}
\bibitem{Dong:arxiv12} Z. Dong, M. Huo, J. Li, J. Li, P. Li, H. Sun, Y. Lu, M. Wang, Y. Wang, and Z. Chen, Visualization of Oxygen Vacancies and Self-doped Ligand Holes in  La$_3$Ni$_2$O$_{7-\delta}$ \href{https://doi.org/10.1038/s41586-024-07482-1}{Nature \textbf{630} 847 (2024).}
\bibitem{Xie:arxiv2024} T. Xie, M. Huo, X. Ni, F. Shen, X. Huang, H. Sun, H. C. Walker, D. Adroja, D. Yu, B. Shen, L. He, K. Cao, and M. Wang, Neutron Scattering Studies on the High-$T_c$ Superconductor  La$_3$Ni$_2$O$_{7-\delta}$ at Ambient Pressure \href{https://doi.org/10.48550/arXiv.2401.12635}{arXiv 2401.12635 (2024).}
\bibitem{Chen:arxiv2024} X. Chen, J. Choi, Z. Jiang, J. Mei, K. Jiang, J. Li, S. Agrestini, M. Garcia-Fernandez, X. Huang, H. Sun, D. Shen, M. Wang, J. Hu, Y. Lu, K.-J. Zhou, and D. Feng, Electronic and magnetic excitations in La$_3$Ni$_2$O$_7$ \href{https://doi.org/10.48550/arXiv.2401.12657}{arXiv 2401.12657 (2024).}
\bibitem{Dan:arxiv2024} Z. Dan, Y. Zhou, M. Huo, Y. Wang, L. Nie, M. Wang, T. Wu, and X. Chen, Spin-density-wave transition in double-layer nickelate La$_3$Ni$_2$O$_7$ \href{https://doi.org/10.48550/arXiv.2402.03952}{arXiv 2402.03952 (2024).}
\bibitem{Takegami:prb} D. Takegami, K. Fujinuma, R. Nakamura, M. Yoshimura, K.-D. Tsuei, G. Wang, N. N. Wang,
J.-G. Cheng, Y. Uwatoko, and T. Mizokawa, Absence of Ni$^{2+}$/Ni$^{3+}$ charge disproportionation and possible roles of O $2p$ holes in La$_3$Ni$_2$O$_{7-\delta}$ revealed by hard x-ray photoemission spectroscopy \href{https://doi.org/10.1103/PhysRevB.109.125119}{Phys. Rev. B \textbf{109}, 125119 (2024).}
\bibitem{Luo:prl23} Z. Luo, X. Hu, M. Wang, W. Wu, and D.-X. Yao, Bilayer Two-Orbital Model of La$_3$Ni$_2$O$_7$ under Pressure \href{https://doi.org/10.1103/PhysRevLett.131.126001}{Phys. Rev. Lett. \textbf{131}, 126001 (2023).}
\bibitem{Zhang:prb23} Y. Zhang, L.-F. Lin, A. Moreo, and E. Dagotto, Electronic structure, dimer physics, orbital-selective behavior, and magnetic tendencies in the bilayer nickelate superconductor La$_3$Ni$_2$O$_7$ under pressure \href{https://doi.org/10.1103/PhysRevB.108.L180510}{Phys. Rev. B \textbf{108}, L180510 (2023).}
\bibitem{Christiansson:prl23} V. Christiansson, F. Petocchi and P. Werner, Correlated Electronic Structure of La$_3$Ni$_2$O$_7$ under Pressure \href{https://doi.org/10.1103/PhysRevLett.131.206501}{Phys. Rev. Lett. \textbf{131},  206501 (2023).}
\bibitem{Yang:prb23} Q.-G. Yang, D. Wang, and Q.-H. Wang, Possible $s_{\pm}$-wave superconductivity in La$_3$Ni$_2$O$_7$ \href{https://doi.org/10.1103/PhysRevB.108.L140505}{Phys. Rev. B \textbf{108}, L140505 (2023).}
\bibitem{Shen:cpl} Y. Shen, M. Qin, and G.-M. Zhang, Effective Bi-Layer Model Hamiltonian and Density-Matrix Renormalization Group Study for the High-$T_c$ Superconductivity in La$_3$Ni$_2$O$_7$ under High Pressure \href{https://doi.org/10.1088/0256-307X/40/12/127401}{Chinese Phys. Lett. \textbf{40}, 127401 (2023).}
\bibitem{Liu:prl23} Y.-B. Liu, J.-W. Mei, F. Ye, W.-Q. Chen, and F. Yang, $s^{\pm}$-Wave Pairing and the Destructive Role of Apical-Oxygen Deficiencies in La$_3$Ni$_2$O$_7$ under Pressure \href{https://doi.org/10.1103/PhysRevLett.131.236002}{Phys. Rev. Lett. \textbf{131}, 236002 (2023).}
\bibitem{Zhang:arxiv1} Y. Zhang, L.-F. Lin, A. Moreo, T. A. Maier, and E. Dagotto, Structural phase transition, $s_{\pm}$-wave pairing, and magnetic stripe order in bilayered superconductor La$_3$Ni$_2$O$_7$ under pressure \href{https://doi.org/10.1038/s41467-024-46622-z}{Nat. Commun. \textbf{15}, 2470 (2024).}
\bibitem{Yang:prb} Y.-F. Yang, G.-M. Zhang, and F.-C. Zhang, Interlayer valence bonds and two-component theory for high-$T_c$ superconductivity of La$_3$Ni$_2$O$_7$ under pressure \href{https://doi.org/10.1103/PhysRevB.108.L201108}{Phys. Rev. B \textbf{108}, L201108 (2023).}
\bibitem{Oh:prb23} H. Oh and Y. H. Zhang, Type-II $t-J$ model and shared superexchange coupling from Hund's rule in superconducting La$_3$Ni$_2$O$_7$ \href{https://doi.org/10.1103/PhysRevB.108.174511}{Phys. Rev. B \textbf{108}, 174511 (2023).}
\bibitem{Liao:prb23} Z. Liao, L. Chen, G. Duan, Y. Wang, C. Liu, R. Yu, and Q. Si, Electron correlations and superconductivity in La$_3$Ni$_2$O$_7$ under pressure tuning \href{https://doi.org/10.1103/PhysRevB.108.214522}{Phys. Rev. B \textbf{108}, 214522 (2023).}
\bibitem{Cao:prb23} Y. Cao, and Y.-F. Yang,  Flat bands promoted by Hund's rule coupling in the candidate double-layer high-temperature superconductor La$_3$Ni$_2$O$_7$ under high pressure \href{https://doi.org/10.1103/PhysRevB.109.L081105}{Phys. Rev. B \textbf{109}, L081105 (2024).}
\bibitem{Lechermann:prb23} F. Lechermann, J. Gondolf, S. B\"otzel, and I. M. Eremin, Electronic correlations and superconducting instability in La$_3$Ni$_2$O$_7$ under high pressure \href{https://doi.org/10.1103/PhysRevB.108.L201121}{Phys. Rev. B \textbf{108}, L201121 (2023).}
\bibitem{Shilenko:prb23} D. A. Shilenko, and I. V. Leonov, Correlated electronic structure, orbital-selective behavior, and magnetic correlations in double-layer La$_3$Ni$_2$O$_7$ under pressure \href{https://doi.org/10.1103/PhysRevB.108.125105}{Phys. Rev. B \textbf{108}, 125105 (2023).}
\bibitem{Jiang:cpl24} K. Jiang, Z. Wang, and F. Zhang, High-Temperature Superconductivity in La$_3$Ni$_2$O$_7$ \href{https://doi.org/10.1088/0256-307X/41/1/017402}{Chin. Phys. Lett. \textbf{41}, 017402 (2024).}
\bibitem{Huang:prb23} J. Huang, Z. D. Wang, and T. Zhou, Impurity and vortex states in the bilayer high-temperature superconductor La$_3$Ni$_2$O$_7$ \href{https://doi.org/10.1103/PhysRevB.108.174501}{Phys. Rev. B \textbf{108}, 174501 (2023).}
\bibitem{Zhang:prb23-2} Y. Zhang, L.-F. Lin, A. Moreo, T. A. Maier, and E. Dagotto, Trends in electronic structures and $s_{\pm}$-wave pairing for the rare-earth series in bilayer nickelate superconductor $R_3$Ni$_2$O$_7$ \href{https://doi.org/10.1103/PhysRevB.108.165141}{Phys. Rev. B \textbf{108}, 165141 (2023).}
\bibitem{Qin:prb23} Q. Qin, and Y.-F. Yang, High-$T_c$ superconductivity by mobilizing local spin singlets and possible route to higher $T_c$ in pressurized La$_3$Ni$_2$O$_7$ \href{https://doi.org/10.1103/PhysRevB.108.L140504}{Phys. Rev. B \textbf{108}, L140504 (2023).}
\bibitem{Zhang:prb24} Y. Zhang, L.-F. Lin, A. Moreo, T. A. Maier, and E. Dagotto, Electronic structure, magnetic correlations, and superconducting pairing in the reduced Ruddlesden-Popper bilayer La$_3$Ni$_2$O$_6$  under pressure: Different role of $d_{3z^2-r^2}$ orbital compared with La$_3$Ni$_2$O$_7$ \href{https://doi.org/10.1103/PhysRevB.109.045151}{Phys. Rev. B \textbf{109}, 045151 (2024).}
\bibitem{Kaneko:prb24} T. Kaneko, H. Sakakibara, M. Ochi, and K. Kuroki, Pair correlations in the two-orbital Hubbard ladder: Implications for superconductivity in the bilayer nickelat La$_3$Ni$_2$O$_7$ \href{https://doi.org/10.1103/PhysRevB.109.045154}{Phys. Rev. B \textbf{109}, 045154 (2024).}
\bibitem{Qu:prl} X.-Z. Qu, D.-W. Qu, J. Chen, C. Wu, F. Yang, W. Li, and G. Su, Bilayer $t-J-J_{\perp}$ Model and Magnetically Mediated Pairing in the Pressurized Nickelate La$_3$Ni$_2$O$_7$ \href{https://doi.org/10.1103/PhysRevLett.132.036502}{Phys. Rev. Lett. \textbf{132}, 036502 (2024).}
\bibitem{Lu:prl} C. Lu, Z. Pan, F. Yang, and C. Wu, Interlayer-Coupling-Driven High-Temperature Superconductivity in La$_3$Ni$_2$O$_7$ under Pressure \href{https://doi.org/10.1103/PhysRevLett.132.146002}{Phys. Rev. Lett. \textbf{132}, 146002 (2024).}
\bibitem{Li:arxiv24} J. Li, P. Ma, H. Zhang, X. Huang, C. Huang, M. Huo, D. Hu, Z. Dong, C. He, J. Liao, X. Chen, T. Xie, H. Sun, M. Wang, Pressure-driven right-triangle shape superconductivity in bilayer nickelate La$_3$Ni$_2$O$_7$ \href{https://doi.org/10.48550/arXiv.2404.11369}{arXiv 2404.11369 (2024).}
\bibitem{Chen:jacs} X. Chen, J. Zhang, A.S. Thind, S. Sharma, H. LaBollita, G. Peterson, H. Zheng, D. Phelan, A. S. Botana, R. F. Klie, and J. F. Mitchell, Polymorphism in the Ruddlesden-Popper Nickelate La$_3$Ni$_2$O$_7$: Discovery of a Hidden Phase with Distinctive Layer Stacking \href{https://doi.org/10.1021/jacs.3c14052}{J. Am. Chem. Soc. \textbf{146}, 23640 (2024).}
\bibitem{Puphal:arxiv12} P. Puphal, P. Reiss, N. Enderlein, Y.-M. Wu, G. Khaliullin, V. Sundaramurthy, T. Priessnitz, M. Knauft, L. Richter, M. Isobe, P. A. van Aken, H. Takagi, B. Keimer, Y. E. Suyolcu, B. Wehinger, P. Hansmann, and M. Hepting, Unconventional crystal structure of the high-pressure superconductor La$_3$Ni$_2$O$_7$ \href{https://doi.org/10.48550/arXiv.2312.07341}{arXiv 2312.07341 (2023).}
\bibitem{Wang:ic} H. Wang, L. Chen, A. Rutherford, H. Zhou, W. Xie, Long-Range Structural Order in a Hidden Phase of Ruddlesden Popper Bilayer Nickelate La$_3$Ni$_2$O$_7$ \href{https://doi.org/10.1021/acs.inorgchem.3c04474}{Inorg. Chem. \textbf{63}, 5020 (2024).}
\bibitem{Abadi:arxiv24} S. N. Abadi, K.-J. Xu, E. G. Lomeli, P. Puphal, M. Isobe, Y. Zhong, A. V. Fedorov, S.-K. Mo, M. Hashimoto, D.-H. Lu, B. Moritz, B. Keimer, T. P. Devereaux, M. Hepting, Z.-X. Shen, Electronic structure of the alternating monolayer-trilayer phase of La$_3$Ni$_2$O$_7$ \href{https://doi.org/10.48550/arXiv.2402.07143}{arXiv 2402.07143 (2024).}
\bibitem{Lechermann:1313} F. Lechermann, S. B\"otze, and Ilya M. Eremin, Electronic instability, layer selectivity and Fermi arcs in La$_3$Ni$_2$O$_7$ \href{https://doi.org/10.48550/arXiv.2403.12831}{arXiv 2403.12831 (2024).}
\bibitem{Zhang:nc20} J. Zhang, D. Phelan, A. S. Botana, Y.-S. Chen, H. Zheng, M. Krogstad, S. G. Wang, Y. Qiu, J. A. Rodriguez-Rivera, R. Osborn, S. Rosenkranz, M. R. Norman, and J. F. Mitchell, Intertwined density waves in a metallic nickelate \href{https://doi.org/10.1038/s41467-020-19836-0}{Nat. Commun.  \textbf{11}, 6003 (2020).}
\bibitem{Samarakoon:prx} A. M. Samarakoon, J. Strempfer, J. Zhang, F. Ye, Y. Qiu, J.-W. Kim, H. Zheng, S. Rosenkranz, M. R. Norman, J. F. Mitchell, and D. Phelan, Bootstrapped Dimensional Crossover of a Spin Density Wave \href{https://doi.org/10.1103/PhysRevX.13.041018}{Phys. Rev. X  \textbf{13}, 041018 (2023).}
\bibitem{Sakakibara:arxiv09} H. Sakakibara, M. Ochi, H. Nagata, Y. Ueki, H. Sakurai, R. Matsumoto, K. Terashima, K. Hirose, H. Ohta, M. Kato, Y. Takano, and K. Kuroki, Theoretical analysis on the possibility of superconductivity in the trilayer Ruddlesden-Popper nickelate La$_4$Ni$_3$O$_{10}$ under pressure and its experimental examination: Comparison with La$_3$Ni$_2$O$_7$ \href{https://doi.org/10.1103/PhysRevB.109.144511}{Phys. Rev. B \textbf{109} 144511 (2024).}
\bibitem{Li:cpl} Q. Li, Y.-J. Zhang, Z.-N. Xiang, Y. Zhang, X. Zhu and H.-H. Wen, Signature of Superconductivity in Pressurized La$_4$Ni$_3$O$_{10}$ \href{https://doi.org/10.1088/0256-307X/41/1/017401}{Chinese Phys. Lett. \textbf{41}, 017401 (2024).}
\bibitem{Zhu:arxiv11} Y. Zhu, E. Zhang, B. Pan, X. Chen, D. Peng, L. Chen, H. Ren, F. Liu, N. Li, Z. Xing, J. Han, J. Wang, D. Jia, H. Wo, Y. Gu, Y. Gu, L. Ji, W. Wang, H. Gou, Y. Shen, T. Ying, X. Chen, W. Yang, C. Zheng, Q. Zeng, J.-G. Guo, and J. Zhao, Superconductivity in trilayer nickelate La$_4$Ni$_3$O$_{10}$ single crystals \href{https://doi.org/10.48550/arXiv.2311.07353}{arXiv 2311.07353 (2023).}
\bibitem{Zhang:arxiv11} M. Zhang, C. Pei, X. Du, Y. Cao, Q. Wang, J. Wu, Y. Li, Y. Zhao, C. Li, W. Cao, S. Zhu, Q. Zhang, N. Yu, P. Cheng, J. Zhao, Y. Chen, H. Guo, L. Yang, and Y. Qi, Superconductivity in trilayer nickelate La$_4$Ni$_3$O$_{10}$ under pressure \href{https://doi.org/10.48550/arXiv.2311.07423}{arXiv 2311.07423 (2023).}
\bibitem{Li:arxiv11} J. Li, C. Chen, C. Huang, Y. Han, M. Huo, X. Huang, P. Ma, Z. Qiu, J. Chen, X. Hu, L. Chen, T. Xie, B. Shen, H. Sun, D. Yao, and M. Wang, Structural transition, electric transport, and electronic structures in the compressed trilayer nickelate La$_4$Ni$_3$O$_{10}$ \href{https://doi.org/10.48550/arXiv.2311.16763}{arXiv 2311.16763 (2023).}
\bibitem{Streltsovt:prb14} S. V. Streltsov and D. I. Khomskii, Orbital-dependent singlet dimers and orbital-selective Peierls transitions in transition-metal compounds \href{https://doi.org/10.1103/PhysRevB.89.161112}{Phys. Rev. B {\bf 89}, 161112(R) (2014).}
\bibitem{Zhang:ossp} Y. Zhang, L. F. Lin, A. Moreo, and E. Dagotto, Prediction of $s^{\pm}$-wave superconductivity enhanced by electronic doping in trilayer nickelates La$_4$Ni$_3$O$_{10}$ under pressure \href{https://doi.org/10.1103/PhysRevB.104.L060102}{Phys. Rev. B \textbf{104}, L060102 (2021).}
\bibitem{Zhang:arxiv24} Y. Zhang, L.-F. Lin, A. Moreo, T. A. Maier, and E. Dagotto, \href{https://doi.org/10.48550/arXiv.2402.05285}{arXiv 2402.05285 (2024).}
\bibitem{Kresse:Prb} G. Kresse and J. Hafner, Ab initio molecular dynamics for liquid metals \href{https://doi.org/10.1103/PhysRevB.47.558}{Phys. Rev. B \textbf{47}, 558 (1993).}
\bibitem{Kresse:Prb96} G.~Kresse and J.~Furthm\"{u}ller, Generalized Gradient Approximation Made Simple \href{https://doi.org/10.1103/PhysRevB.54.11169}{Phys. Rev. B \textbf{54}, 11169 (1996).}
\bibitem{Blochl:Prb} P. E. Bl\"{o}chl, Projector augmented-wave method \href{https://doi.org/10.1103/PhysRevB.50.17953}{Phys. Rev. B \textbf{50}, 17953 (1994).}
\bibitem{Perdew:Prl} J. P. Perdew, K. Burke, and M. Ernzerhof, Generalized Gradient Approximation Made Simple \href{https://doi.org/10.1103/PhysRevLett.77.3865}{Phys. Rev. Lett. \textbf{77}, 3865 (1996).}
\bibitem{Momma:vesta} K. Momma and F. Izumi, VESTA 3 for three-dimensional visualization of crystal, volumetric and morphology data \href{https://doi.org/10.1107/S0021889811038970}{J. Appl. Crystallogr. \textbf{44}, 1272 (2011).}
\bibitem{HP}{Here, based on the lattice structure of the P4/mmm phase at 16 GPa~\cite{Puphal:arxiv12}, we fully relaxed the atomic positions until the Hellman-Feynman force on each atom was smaller than $0.001$ eV/{\AA}.}
\bibitem{U4}{In addition, we also studied the electronic structures by introducing additional Hubbard $U$ effects by using the Dudarev's rotationally invariant formulation~\cite{Dudarev:prb}, where the main conclusion does not change~\cite{Supplemental}).}
\bibitem{Dudarev:prb} S. L. Dudarev, G. A. Botton, S. Y. Savrasov, C. J. Humphreys, and A. P. Sutton, Electron-energy-loss spectra and the structural stability of nickel oxide:  An LSDA+U study \href{https://doi.org/10.1103/PhysRevB.57.1505}{Phys. Rev. B \textbf{57}, 1505 (1998).}
\bibitem{Supplemental}{See Supplemental Material for method details and more results.}
\bibitem{Zhang:prb20} Y. Zhang, L.-F. Lin, W. Hu, A. Moreo, S. Dong, and E. Dagotto, Similarities and differences between nickelate and cuprate films grown on a SrTiO$_3$ substrate \href{https://doi.org/10.1103/PhysRevB.102.195117}{Phys. Rev. B \textbf{102}, 195117 (2020).}
\bibitem{Mostofi:cpc} A. A. Mostofi, J. R. Yates, Y. S. Lee, I. Souza, D. Vanderbilt, and N. Marzari, wannier90: A tool for obtaining maximally-localised Wannier functions \href{https://doi.org/10.1016/j.cpc.2007.11.016}{Comput. Phys. Commun. \textbf{178}, 685 (2007).}
\bibitem{Kubo2007} K. Kubo, Pairing symmetry in a two-orbital Hubbard model on a square lattice \href{https://doi.org/10.1103/PhysRevB.75.224509}{Phys. Rev. B \textbf{75}, 224509 (2007).}
\bibitem{Graser2009} S. Graser, T. A. Maier, P. J. Hirschfeld, and D. J.  Scalapino, Near-degeneracy of several pairing channels in multiorbital models for the Fe pnictides \href{https://doi.org/10.1088/1367-2630/11/2/025016}{New J. Phys. \textbf{11}, 25016 (2009).}
\bibitem{Altmeyer2016} M. Altmeyer, D. Guterding, P. J. Hirschfeld, T. A. Maier, R. Valent\'{\i}, and D. J. Scalapino, Role of vertex corrections in the matrix formulation of the random phase approximation for the multiorbital Hubbard model \href{https://doi.org/10.1103/PhysRevB.94.214515}{Phys. Rev. B \textbf{94}, 214515 (2016).}
\bibitem{Romer2020} A. T. R{\o}mer, T. A. Maier, A. Kreisel, I. Eremin, P. J. Hirschfeld, and B. M. Andersen, Pairing in the two-dimensional Hubbard model from weak to strong coupling \href{https://doi.org/10.1103/PhysRevResearch.2.013108}{Phys. Rev. Res. \textbf{2}, 013108 (2020).}
\bibitem{Tian:prb24} Y.-H. Tian, Y. Chen, J.-M. Wang, R.-Q. He, and Z.-Y. Lu, Correlation effects and concomitant two-orbital $s_{\pm}$-wave superconductivity in La$_3$Ni$_2$O$_7$ under high pressure \href{https://doi.org/10.1103/PhysRevB.109.165154}{Phys. Rev. B \textbf{109}, 165154 (2024).}

\end{references}
\end{document}